\documentstyle[12pt]{article}
\begin{document}
\begin{center}
\large
{THE CHIRAL OSCILLATOR AND ITS APPLICATIONS IN QUANTUM THEORY}
\end{center}
\vskip 2cm
R. Banerjee\footnote{E-mail:rabin@boson.bose.res.in}\\
S. N. Bose National Centre for Basic Sciences\\
Block JD, Sector III, Calcutta 700091, India\\

\vskip .5cm
\noindent and\\
\vskip .5cm
\noindent Subir Ghosh\\
Dinabandhu Andrews College\\
Garia, West Bengal, India.\\
\vskip 2cm
\noindent Abstract\\
The fundamental importance of the chiral oscillator is elaborated. Its quantum
invariants are computed. As an application the Zeeman effect is analysed.
We also show that the chiral oscillator is the most basic 
example of a duality invariant model, simulating the effect of the familiar
electric-magnetic duality.
\newpage
It is well known that the Harmonic Oscillator (HO) pervades our understanding
of quantum mechanical as well as field theoretical
models in various contexts. An interesting thrust in
this direction was 
recently made in \cite{rprl} where the quantum invariants of the HO
were computed. It was also opined that this approach could be used for 
developing a technique \cite{rjmp} to study interacting and time dependent 
(open) systems.

In this paper we argue that, in some instances, the Chiral Oscillator (CO)
instead of the usual HO captures the essential physics of the problem. This
is tied to the fact 
that the CO simulates the left-right symmetry. Consequently the CO
has a decisive role in those cases where this symmetry is significant.

The CO is first systematically derived from the HO and the issue of
symmetries is clarified. Indeed, it is explicitly shown that the
decomposition of the HO leads to a pair of left-right symmetric CO's. The
soldering of these oscillators to reobtain the HO is an instructive
exercise. Following the methods of \cite{rprl,rjmp}, the quantum invariants
of the CO's are computed and their connection with the HO invariant is
illuminated. As an application, the Zeeman splitting \cite{pk} for the
Hydrogen atom
electron energy levels under the influence of a constant magnetic field is
studied. The interaction of the atom with a time-dependent magnetic field,
constituting an open system, can also be analysed from the general expressions.
In a completely different setting we show that the CO is 
the most basic example of a
duality invariant theory 
\cite{az}.  By reexpressing the computations in a
suggestive electromagnetic notation, the mapping 
of this duality with Maxwell's 
electromagnetic duality is clearly established.

The Lagrangean for the one dimensional HO is given by
\begin{equation}
L={M\over 2}(\dot {x}^2-\omega^2x^2).
\label{eqlho}
\end{equation}
To obtain the CO, the basic step is to convert (\ref{eqlho}) in a first order
form by introducing an auxiliary variable $\Lambda$ in a symmetrised form,
\begin{equation}
L={M\over 2}(\Lambda\dot x-x\dot{\Lambda}-{\Lambda^2}-\omega^2x^2).
\label{eqlf}
\end{equation}
There are now two distinct classes for relabelling these variables 
corresponding to proper and improper rotations generated by the matrices
with determinant $\pm1$,
\[ \left (
\begin{array}{c}
x\\
{{\Lambda}\over{\omega}}\\
\end{array}
\right )=\left (
\begin{array}{cc}
cos\theta & sin\theta\\
-sin\theta & cos\theta
\end{array}
\right )\left (
\begin{array}{c}
x_1 \\
x_2\\
\end{array}\right ),
~~ \left (
\begin{array}{c}
x\\
{{\Lambda}\over{\omega}}\\
\end{array}
\right )=\left (
\begin{array}{cc}
sin\phi & cos\phi\\
cos\phi & -sin\phi
\end{array}
\right )\left (
\begin{array}{c}
x_1 \\
x_2\\
\end{array}\right )
\]
leading to the structures,
\begin{equation}
L_{\pm}={M\over 2}(\pm\omega\epsilon_{\alpha\beta}x_\alpha \dot{x}_\beta
-\omega^2x_\alpha^2),
\label{eqlco}
\end{equation}                                                                     
where $\alpha=1,2$ is an internal index with $\epsilon_{12}=1$. 
The basic Poisson brackets
of the above model are read off from the symplectic structure,
\begin{equation}
\{x_\alpha ,x_\beta \}_{\pm}=\mp{1\over{\omega M}}\epsilon_{\alpha\beta}.
\label{eqbr}
\end{equation}
The corresponding Hamiltonians are,
\begin{equation}
H_{\pm}={{M\omega^2}\over 2}(x_1^2+x_2^2) =\tilde{H}.
\label{eqhco}
\end{equation}

The above Lagrangeans in (\ref{eqlco}) are interpreted as
two bi-dimensional CO's rotating 
in either a clockwise or an anti-clockwise sense. A
simple way to verify this property is to look at the spectrum of the
angular momentum operator,
\begin{equation}
\omega J_{\pm}=\omega\epsilon_{\alpha\beta}
x_\alpha p_\beta=\pm{1\over 2}M\omega^2x_\alpha^2
=\pm\tilde{H},
\label{eqJ}
\end{equation}
where $\tilde{H}$ is defined above.  

To complete the picture it is desirable to show the mechanism of
combining the left and right CO's to reproduce the usual HO. This is achieved
by the soldering technique \cite{adc,abc} introduced recently. Let us then begin with two {\it independent} chiral Lagrangeans
$L_+(x)$ and $L_-(y)$. Consider
the following gauge transforms,
$\delta x_\alpha=\delta y_\alpha =\eta_\alpha$ under which
$$
\delta L_{\pm}(z)=M\omega\epsilon_{\alpha\beta}\eta_\alpha 
(\pm\dot{z}_\alpha
+\omega\epsilon_{\alpha\beta}z_\beta),~~~z=x,y.
$$
Introduce a new variable $B_\alpha$, which will effect the soldering,
transforming as,
$\delta B_\alpha =\epsilon_{\alpha\beta}\eta_\beta$. This new Lagrangean
\begin{equation}
L=L_{+}(x)+L_{-}(y)-M\omega B_\alpha(\dot{x}_\alpha
+\omega\epsilon_{\alpha\beta}x_\beta-\dot{y}_\alpha
+\omega\epsilon_{\alpha\beta}y_\beta),
\label{eqlinv}
\end{equation}
is invariant under the above transformations.
Eliminating $B_\alpha$ by the equations of motion, we obtain the final soldered
Lagrangean,
$$L(w)={M\over 4}(\dot{w}^2_{\alpha}-\omega^2 w^2_\alpha),$$
which is no longer a function of $x$ and $y$ independently, but only on
their gauge invariant combination, $w_\alpha=x_\alpha -y_\alpha$. The
soldered Lagrangean just corresponds to a bi-dimensional 
simple harmonic oscillator.
Thus, by starting from two distinct Lagrangeans containing the opposite
aspects of chiral symmetry, it is feasible to combine them into a single 
Lagrangean.

The connection between the CO and HO is now used to obtain the invariants
of the former by exploiting known results \cite{rprl} for the latter. 
 For the positive CO, 
\begin{equation}
I^+={1\over 2}\tan^{-1}({x_1}^{-1}x_2)+ {1\over 2}\tan^{-1}(x_2{x_1}^{-1}),
\label{eqco+}
\end{equation}
is the invariant, while $I^-$ is given by interchanging $x_1$ and $x_2$. Note 
that non-commutativity of the variables has already been taken into
account. Incorporating the "soldering" prescription 
\cite {abc} whereby we were able
to construct a bi-dimensional oscillator 
from the two CO's, another
quantum invariant can also be obtained,
\begin{equation}
I^+(x_1,x_2) \oplus I^-(y_1,y_2)=I(x_1-y_1,x_2-y_2),
\label{eqsol}
\end{equation}
where, the right hand side of the equation is a simple sum of two terms,
obtained by
substituting $x_1-y_1,~~M(\dot x_1-\dot y_1)/2$ and
 $x_2-y_2,~~M(\dot x_2-\dot y_2)/2$
in place of $x$ and $p$ in the corresponding expression for HO in \cite{rprl}.
 We stress that the above 
invariant operators are independent as they pertain to completely
different systems and were not present in the literature so far. In the next
section, we will put the CO invariants into direct use in interacting and
open quantum systems by considering the Zeeman effect.

Let us consider the simplistic Bohr model of Hydrogen atom, where the
(non-relativistic) electron is moving in the presence of a repulsive
centrifugal barrier and the attractive Coulomb potential. The effective
central potential has a well like structure and we consider the 
standard HO approximation about the potential minimum. The excitations are
the HO states above the minimum. Hence the electron, at a particular
stationary state, is approximated to an oscillator with a 
frequency
$\omega$, obtained from the effective potential 
seen by the atomic electron without
the magnetic field.This
yields $\omega=(Me^4)/l^3$, with $l=Mr^2\dot{\phi}$
being the angular momentum, when expressed in plane polar coordinates.

In the presence of a magnetic field ${\bf B}$, the motion of the electron
can be broken into components parallel and perpendicular to ${\bf B}$. The
Lorentz force acting on the electron affects the motion in the normal plane of
${\bf B}$ only, the motion being two rotational modes in the clockwise
and anti-clockwise sense, or more succintly two CO's of opposite chirality.
In this setup, ${\bf B}$ splits the 
original level into three levels, one of
frequency $\omega$ remaining unchanged and the other two frequencies
changed to $\omega\pm(eB)/(2Mc)$ \cite{pk}. This clearly shows that there is a 
redundancy in the number of degrees of freedom in treating the electron
as a HO, whereas the CO representation is more elegant and economical
whenever the degeneracy between the right and left movers is lifted such as
in the presence of magnetic field.

The Hamiltonian of a charged HO in an axially symmetric magnetic field is,
$${\bf A}={1\over 2}B(t){\bf k}\times{\bf r},~~{\bf B}(t)=\nabla\times{\bf A}
=B(t){\bf k}$$
$$H={1\over{2M}}({\bf p}-e{\bf A})^2 +{1\over 2}M\omega^2r^2
={1\over{2M}}({p_1}^2+{p_2}^2)+{1\over 2}M\omega^2({x_1}^2+{x_2}^2)$$
\begin{equation}
+{{eB(t)}\over{2Mc}}(x_2p_1-x_1p_2)
+{{e^2}\over{8Mc^2}}{B(t)}^2({x_1}^2+{x_2}^2).
\label{eqcho}
\end{equation}

For the semi-classical reasoning (regarding the Zeeman effect) to hold,
$\mid{\bf B}\mid$ must be small in the sense that the radius
of gyration $r=~(cMv)/(eB)=~(cl)/(eBr)$, which simplifies to
$r=~\sqrt{(cl)/(eB)}=~\sqrt{(nc\hbar)/(eB)}$
is much larger than the Bohr radius of the (Hydrogen) atom \cite{yk}
$r_{Bohr}=~\hbar^2/(Me^2).$
This condition is expressed as
\begin{equation}
{{\hbar^3B}\over{cM^2e^3}}<<1.
\label{eqsb}
\end{equation}
In our Hamiltonian (\ref{eqcho}), this condition will hold if
\begin{equation}
\mid{1\over 2}M\omega^2({x_1}^2+{x_2}^2)\mid>>
\mid{{eB(t)}\over{2Mc}}(x_2p_1-x_1p_2)\mid.
\label{eqsm}
\end{equation}
To verify this, substitute $\omega=~Me^4/l^3 $ and $({x_1}^2+{x_2}^2)=~r_
{Bohr}$ in the left hand side, and $(x_2p_1-x_1p_2)=~l$ in the right hand
side. This reproduces (\ref{eqsb}). The quadratic
$B$-term in (\ref{eqcho}) is still smaller.

The above structure of the Hamiltonian is very similar 
to the model of a charged particle in a
specified electromagnetic field, considered in \cite{rprl,
rjmp}. The idea there is
to look for the invariants of the full interacting Hamiltonian, 
and to construct eigenstates of the invariant operator. The
solutions of the time dependent Schrodinger equation are related uniquely
to these eigenstates via a time dependent phase,
$$
\mid\lambda,k,t>_{Sch}=e^{i\alpha_{\lambda k}(t)}\mid\lambda,k,t>_{I},~~~~
I(t)\mid\lambda,k,t>_{I}=\lambda \mid\lambda,k,t>_{I},$$
satisfying,
$$i\hbar{{d\alpha_{\lambda k}}\over{dt}}=<\lambda,k\mid_I(i\hbar{{\partial}
\over{\partial t}}-H)\mid\lambda,k>_I.$$
Next we define,
$$H_o={1\over{2M}}({p_1}^2+{p_2}^2)+{1\over 2}M\omega^2({x_1}^2+{x_2}^2)$$
and the rest of the $B$-dependent terms 
appearing in (\ref{eqcho}) as small perturbations. In the framework
of \cite{rpla}, the invariant 
operator is also expressible as a power series in the small 
parameter $B\hbar^3/(cM^2e^3)$
 and the zeroth order invariant $I_0$ is identical 
to $H_o$. Hence the
eigenstates of $H_o$ and $I_o$ will be same and $\mid\lambda,k,t>_I=
exp(-i(n+{1\over 2})\omega t)\mid\lambda,k>_I$. As in the conventional
scenario, the total energy is also expressed as a series with the zeroth term
being $(n+{1\over 2})\hbar\omega$. Thus we will compute the $B$-dependent
corrections only by the scheme of \cite{rjmp}, which actually comprises
the task of calculating the phase $\alpha_{\lambda k}$. Here the CO's will 
come into play. 

As we have already established the connection between the results of
HO and CO models, we simply replace the HO variables by the CO ones in 
the final result. From the symplectric structure, the following
identifications are consistent,
\begin{equation}
CO^+:~~\{{x_1}^+,~{x_2}^+\}=-{1\over{\omega M}}
,~~\to p_1\equiv -\omega M{x_2}^+,~p_2\equiv \omega M{x_1}
^+,
\label{px+}
\end{equation}
\begin{equation}
CO^-:~~\{{x_1}^-,~{x_2}^-\}={1\over{\omega M}}
,~~\to p_1\equiv \omega M{x_2}^-,~p_2\equiv -\omega M{x_1}
^-.
\label{px-}
\end{equation}
Introducing these in (\ref{eqcho}), we get,
\begin{equation}
H_{\pm}={M\over 2}({x_1}^2+{x_2}^2)(1+{{e^2B^2}\over
{4M^2c^2}}\mp{{eB}\over{Mc}}).
\label{cco}
\end{equation}
The above splitting in the energy is one of our main results. This underlines
the economy in the CO formulation since one CO is sufficient to obtain the
correct results. Obviously it is easier to work with less number of degrees
of freedom in cases of more complicated systems. Essentially 
this change in the relative sign of the linear $B$ term 
can also be interpreted as a consequence of the
opposite angular momenta of the CO's, as demonstrated before. This brings
us to the cherished expression of the phase for the two CO's,
\begin{equation}
\alpha_{jn}^{\pm}=\mp[n+(j+{1\over 2})]{e\over{Mc}}\int^t dt'[{1\over 2}
B(t')-\rho^{-2}(t')],
\label{eqpco}
\end{equation}
where the quantum numbers $j$ and $n$ are explained in \cite{rjmp} and $\rho
(t')$ satisfies the equation,
$$({{Mc}\over e})^2\ddot{\rho}+{{B^2(t)}\over 2}\rho -\rho^{-3}=0.$$
Considering the simplest case, that is normal Zeeman effect, where $B$ is a
constant, we find a time-independent solution of $\rho$, 
$\rho^2=\pm{\sqrt 2}/B$.
When $\rho^2=-{\sqrt 2}/B$ is
substituted in (\ref{eqpco}), the standard
Zeeman level splitting is reproduced.
\begin{equation}
E_n^{\pm}=(n+{1\over 2})\hbar\omega\pm[n+(j+{1\over 2})]{{eB}\over{Mc}}.
\label{eqzee}
\end{equation}
On the other hand, $\rho^2={\sqrt 2}/B$ reveals no shift in the
energy eigenvalue. Clearly this is reminiscent of the fact that the
energy of the mode parallel to ${\bf B}$ remains unaffected.
For time dependent magnetic field, one has to obtain the appropriate solution 
for $\rho$. Inserting this in (\ref{eqpco}) it is then possible to obtain
the solutions of the corresponding Schr$\ddot o$dinger equation.

We next show the possibility of interpreting the CO as a prototyype of a
duality invariant model characteristic of the electric-magnetic duality
\cite{az}. For convenience, we set $M=~\omega=~1$ in (\ref{eqlho}). 
Introduce a change of variables,
$E=\dot x,~~~B=x$,
so that
\begin{equation}
\dot B -E=0
\label{eqeb}
\end{equation}
is identically satisfied. In these variables, the Lagrangian (\ref{eqlho}) and 
the corresponding equation of motion  are expressed as
\begin{equation}
L={1\over 2}(E^2-B^2),
~~~\dot E+B=0.
\label{eqem}
\end{equation}
It is simple to observe that the transformations, \footnote{Note that these
are the discrete cases $(\theta=~\pm{{\pi}\over 2})$ for a general $SO(2)$
rotation matrix parametrised by the angle $\theta$.}
$E\rightarrow \pm B;~B\rightarrow\pm E$,
swap the equation of motion
in (\ref{eqem}) with the identity (\ref{eqeb}) although the
Lagrangean (\ref{eqem}) is not invariant. The similarity with the
corresponding analysis in Maxwell theory is quite striking, with $x$ and
$\dot x$ simulating the roles of the magnetic and electric fields,
respectively. There is a duality among the equation of motion and the
"Bianchi" identity (\ref{eqeb}), which is not manifested in the Lagrangean.

Let us now consider the Lagrangean for the CO,
\begin{equation}
L_{\pm}={1\over 2}(\pm\epsilon_{\alpha\beta}x_\alpha\dot{x}_\beta
-x^{2}_\alpha)
={1\over 2}(\pm\epsilon_{\alpha\beta}B_\alpha E_\beta
-B^{2}_\alpha).
\label{eqdu}
\end{equation}
These chiral Lagrangeans are manifestly invariant under the duality
transformations,
\begin{equation}
x_\alpha\rightarrow~R^{+}_{\alpha\beta}(\theta)x_\beta.
\label{eqxr}
\end{equation}
Thus, the CO's represent a quantum
mechanical example of a duality invariant model. Indeed, the expressions for
$L_{\pm}$ given in the second line of (\ref{eqdu}) closely resemble the
analogous structure for the Maxwell theory deduced in \cite{bw}.

The generator of the infinitesimal symmetry transformmation is given by,
$Q=x_\alpha x_\alpha/2,$
so that the complete transformations (\ref{eqxr}) are generated by,
$$x_\alpha\rightarrow~x_{\alpha}'=e^{-i\theta Q}x_\alpha e^{i\theta Q}
=~R^{+}_{\alpha\beta}(\theta)x_\beta.$$
This follows by exploiting the basic bracket of the theory given in 
(\ref{eqbr}).

To conclude, certain interesting properties of the CO were
illustrated. A systematic method of obtaining this oscillator from the
usual simple HO was given. It was also shown that the
distinct left and right components of the CO were combined by the soldering
formalism \cite{adc, abc} to yield a bi-dimensional HO. In this way the
symmetries of the model were highlighted. The importance in the CO lies in
the fact that in some cases it has a concrete and decisive role than the
usual simple HO in illuminating the basic physics of the problem. This was
particularly well seen in the derivation of the Zeeman splitting by
exploiting the perturbation theory technique based on quantum invariant
operators \cite{rprl, rjmp}. An explicit computation of the quantum
invariants for the CO was also performed. Apart from the study of the Zeeman
effect, such CO invariants can find applications in other quantum mechanical
examples, particularly where a left-right symmetry is significant. Another
remarkable feature of the present analysis has been the elucidation of the
fundamental nature of the duality symmetry currently in vogue either in
quantum field theory or the string theory \cite{az, bw}. It was shown that 
the CO was a duality symmetric model, contrary to the usual HO. Expressed
in the "electromagnetic" notation, this difference was seen to be the origin
of the presence or absence of duality symmetry in electrodynamics.

It may be remarked that the explicit demonstration of duality symmetry in
a quantum mechanical world is nontrivial since conventional analysis
\cite{az} considers two types of duality invariance confined to either
$D=4k$ or $D=4k+2$ dimensions, thereby leaving no room for $D=1$ dimension.
Nevertheless, since most field theoretical problems can be understood on
the basis of the HO, it is reassuring to note that the origin of 
electromagnetic duality invariance is also contained in a variant of the
HO- the chiral oscillator. Our study clearly reveals that the CO complements
the usual HO in either understanding or solving various problems in quantum
theory.

\end{document}